\documentclass{PoS}

\title{Intersecting Brane Vacua and Non-perturbative Transitions}

\ShortTitle{Intersecting Brane Vacua and Non-perturbative Transitions}

\author{\speaker{Cezar Condeescu}\thanks{I am very thankful to C. Angelantonj, E. Dudas and G. Pradisi  for an enjoyable collaboration.}\vspace{0.6cm}\\
          Department of Theoretical Physics\\
          Horia Hulubei National Institute of Physics and Nuclear Engineering - IFIN-HH\\
          Atomistilor 407, P.O. Box MG-6, Mãgurele, 077125, jud Ilfov, Rom\^{a}nia\\
          and\\
          Centre de Physique Th\'eorique, Ecole Polytechnique and CNRS\\
          91128 Palaiseau, France\\
          E-mail: \email{ccezar@theory.nipne.ro}}

\abstract{We discuss the process of transmutation of branes into Abelian magnetic flux on the worldvolume of $D9$ branes. This phenomenon is T-dual to the recombination of branes in the intersecting-brane description. In the low energy limit this can be described by a Higgs mechanism involving bifundamental scalars. The Abelian magnetic background arises as a consequence of the compactness of the internal space.
Explicit (six-dimensional) examples based on the supersymmetric and the Brane Supersymmetry Breaking (BSB) $\mathbb{T}^4/\mathbb{Z}_2$ orientifold are analyzed.  }

\FullConference{The 2011 Europhysics Conference on High Energy Physics, EPS-HEP 2011,\\
		July 21-27, 2011\\
		Grenoble, Rh\^one-Alpes, France}

\begin{document}

\section{Introduction}
Compactification of string theory leads to a multitude of vacuum solutions. We consider in the following the possibility of having (non-perturbative)
transitions among different vacua. In the context of type I theory it is well known that $D5$ branes realize gauge instanton effects in the zero-size limit for the theory living on the worldvolume of the $D9$ branes \cite{Witten:1995gx}. By blowing up the instanton size one can connect string orientifold vacua with $D9/D5$ branes to vacua where internal magnetic fluxes are present. The transmutation of D-branes into magnetic flux \cite{Blumenhagen:2000fp,Angelantonj:2000hi,as} is T-dual to brane recombination in the framework of intersecting branes \cite{csu,Angelantonj:2011hs}. In the low energy limit this phenomenon can be captured by a Higgs mechanism. Scalars living at the intersection of branes condense and this condensate interpolates between zero-size instantons and constant non-Abelian magnetic field \cite{Witten:1995gx}. In \cite{Angelantonj:2011hs} the analysis is extended for the case of an Abelian theory on the worldvolume of the $D9$ branes and compact internal spaces. The constant magnetic field is related to the zero-mode of the Laplacian on a compact space. We discuss the examples based on the $\mathbb{T}^2/\mathbb{Z}_2$ including both the supersymmetric and the Brane Supersymmetry Breaking \cite{bsb1a,bsb1b,bsb1c,bsb1d,Angelantonj:1999ms} cases. The conclusion that arises is that all string vacua, in a given orientifold construction, live in the same moduli space and are connected to one another by the process of brane recombination.\\
The organization is as follows. In section 2 we discuss the transition from the point of view of brane recombination and the corresponding Higgs effect. In section 3 we consider a field theory analysis which shows the connection between the {\it vev} of the Higgs and the constant magnetic field.
\section{Brane Recombination and Higgs Effect}
We consider type IIB string theory with intersecting $D7$ branes compactified on $\mathbb{T}^4/\mathbb{Z}_2$ with the usual orientifold projection $\tilde\Omega=\Omega(-1)^{F_L} {\cal R}$ which yields supersymmetric six-dimensional vacua \footnote{For notations and conventions see \cite{Angelantonj:2011hs}.}. The rules for brane recombination essentially reduce to conservation of the RR charges. This is neatly encoded in the description in terms of cycles were the brane recombination process corresponds to the addition of cycles in the homology (K-theory) of the compact internal space. The conservation of RR charges is then automatic. One represents $D7$ branes by two cycles $\Pi_a\in H_2(\mathbb{T}^4/\mathbb{Z}_2)$. The homology of the orbifold space contains, in addition to the cycles inherited from the covering torus called bulk cycles, exceptional cycles associated with each of the orbifold fixed points. However, the orientifold invariant combination $\Pi_a+\tilde\Omega\Pi_a\equiv \Pi_{a}+\Pi_{\bar a}$ only wraps bulk cycles matching the fact that the twisted tadpole conditions are identically vanishing. The conservation of RR charges for the recombination of two stacks of branes $a,b$ into one stack $c$ can be expressed formally as $N_c(\Pi_c+\Pi_{\bar c})= N_a(\Pi_a+\Pi_{\bar a})+N_{b}(\Pi_b+\Pi_{\bar b})$. In terms of the torus wrapping numbers this yields
\begin{equation}
    N_c m_c^1m_c^2=N_a m_a^1m_a^2+N_b m_b^1m_b^2\, , \qquad N_c n_c^1n_c^2=N_a n_a^1n_a^2+N_b n_b^1n_b^2
\end{equation}
Consider the following example with two stacks of branes $\Pi_{D7}\sim (1,0;1,0)$, $\Pi_{D7'}\sim(0,1;0,1)$ based on the $U(16)\times U(16)$ gauge group with the massless spectrum consisting of hypermultiplets in $2\times [(120,1)+(1,120)]$ and $(16,16)$. Recombining all branes yields a model with one stack of branes with homology cycle described by the following wrapping numbers $\Pi_r\sim (1,1;1,1)$ and $U(16)$ gauge group. Its massless spectrum consists of $4$ hypermultiplets in the $120$ representation. It is easy to see that this process is described at the field theory level by a Higgs effect involving a {\it vev} for the bifundamental $(16,16)$. In general one can obtain through recombinations all solutions to the tadpole conditions in the given orientifold background.\\
We now turn to the non-supersymmetric version of the $\mathbb{T}^4/\mathbb{Z}_2$ orientifold. In this case, the projection is modified to $\hat\Omega = \tilde\Omega \sigma$, yielding six-dimensional BSB vacua. One of the two $O7$ planes is exotic (i.e. it has positive NSNS tension and positive RR charge). The orientifold invariant combination $\Pi_{a}+\Pi_{\bar a}= 2c_a^1{\bf \pi}_1+2c_a^2{\bf \pi}_2+2\sum_{x,y}\epsilon_a^{xy} {\bf e}_{xy}$ contains now exceptional cycles in agreement with the fact that twisted tadpole conditions are non-trivial. The recombination of two stacks of branes $a,b$ now reads
\begin{equation}
    (\Pi_a+\Pi_{\bar a})+(\Pi_b+\Pi_{\bar b})=(m^1_am^2_a+m^1_bm^2_b){\bf \pi}_1+(n^1_an^2_a+n^1_bn^2_b){\bf \pi}_2+2\sum_{x,y}(\epsilon_a^{xy}+\epsilon_b^{xy}){\bf e}_{xy}
\end{equation}
A similar example to the supersymmetric one above involves now four stacks of branes wrapping the cycles $\Pi_{D7}^{\pm}\sim (1,0;1,0)$, $\Pi_{D7'}^{\pm}\sim(0,1;0,1)$. The gauge group of the model is $SO(16)^2\times USp(16)^2$. Recombining all branes now yields a two stack model with $U(8)\times U(8)$ gauge group and homology cycles $\Pi_r^{\pm}\sim (1,1;1,1)$ with opposite twisted charges. An analysis of the field theory Higgs effect shows that extra states need to acquire a mass in order to reproduce the correct massless spectrum of the recombined model. The necessary third order Yukawa couplings and fourth order scalar couplings exist. This indicates that also in the non-supersymmetric case brane recombination can connect all vacua on the given orientifold.

\section{Field Theory Solutions for the D5 vs Magnetized D9 transition}
It is instructive to analyze the field theory equations corresponding to the transmutation of branes into constant magnetic flux. Our analysis is in the context of type I theory with $D9/D5$ branes. For illustrative purposes we consider first six-dimensional Super-Yang-Mills theory compactified on a single two torus $\mathbb{T}^2$. We denote by $A_4, A_5$ the components of the gauge field along the two torus and introduce the complex field $\Phi=A_5+iA_4$. The scalar potential of the theory is of the form $V=\frac{1}{2}D^2+|F|^2$. We parametrize the {\it vev} of the $95$ states by a localized FI term in the action
\begin{equation}
  D=-\frac{1}{2}(\partial \Phi +\bar\partial\Phi)+\xi \delta^{(2)}(z)\, , \qquad F=0
  \end{equation}
The equations of motion are very simple $\partial D=\bar\partial D=0$ and one has the following solution
\begin{equation}
\Phi=\xi\ \partial G_2 \qquad D=\frac{\xi}{V}
\end{equation}
Notice above that the D-term is equal to the generated constant magnetic field. Making use of the explicit expression of the Green's function on a two torus one can see that the magnetic field arises from the zero-mode of the Laplacian on a compact space
\begin{equation}
\Phi (z)  =  - \frac{ \alpha \, \theta_1^\prime  (z | \tau )}{4 \, \theta_1 (z | \tau )}  -
\frac{\pi \alpha\, (z- {\bar z})}{4 \, {\rm Im}\, \tau }
\end{equation}
The last term in the equation above gives rise to the magnetic field which is T-dual to the angle of the brane. The situation is similar for compactifications on a four torus $\mathbb{T}^4$ in what concerns the origin of the magnetic field. In this case we take eight-dimensional Super-Yang-Mills theory compactified on $\mathbb{T}^4$ with complex gauge fields $\Phi_1, \Phi_2$ along each of the two tori. The F- and D- terms are given by
\begin{equation}
\bar F = -\frac{1}{\sqrt{2}}\, (\partial_1\Phi_2-\partial_2\Phi_1) \,, \qquad D=-\frac{1}{2}\, \sum_{i=1}^2(\partial_i\bar \Phi^i+\bar\partial^i\Phi_i)+\xi \,\delta^{(4)}\,
\end{equation}
The solution to the equations of motion $\bar\partial^i D - \sqrt{2}\, \epsilon^{ij}\, \partial_j F=0$ can be written as
\begin{equation}
\Phi_1 =  k_1 \, \partial_1 \, G_4 \, , \qquad \Phi_2  =  k_2 \, \partial_2 \, G_4 \,
\end{equation}
With this solution the F- and D- terms become
\begin{eqnarray}
&\bar F=0 \qquad {\rm if}\qquad k_1=k_2=k \nonumber\\
&D=-(F_{45}+F_{67})=\frac{\xi}{V_4} \qquad {\rm with} \qquad k=\xi
\end{eqnarray}
Rewriting the solution in terms of real components one obtains the following
\begin{equation}
A_\mu=M_{\mu\nu}\partial_\nu G_4=M_{\mu\nu}x_\nu f(r) \qquad {\rm with}\qquad M_{\mu\nu}=k\left(
                                                                                                          \begin{array}{cccc}
                                                                                                            0 & -1 & 0 & 0 \\
                                                                                                            1 & 0 & 0 & 0 \\
                                                                                                            0 & 0 & 0 & -1 \\
                                                                                                            0 & 0 & 1 & 0 \\
                                                                                                          \end{array}
                                                                                                        \right)
\end{equation}
The solution above is very similar to the $SU(2)$ instanton solution in the singular gauge. However in the U(1) case that we considered the singularity cannot be gauged away. Due to the fact that we are on a compact space the corresponding field strength is not selfdual $\tilde F_{\mu\nu}=-F_{\mu\nu}-M_{\mu\nu}\Delta G_4$. The singularity in the origin leads to an infinite instanton number \cite{Angelantonj:2011hs}.\\
In conclusion, we have considered non-perturbative transitions among string vacua with intersecting/magnetized branes. By studying specific six-dimensional compactifications based on the supersymmetric and non-supersymmetric (BSB) versions of the $\mathbb{T}^4/\mathbb{Z}_2$ orientifold, we have argued that all vacua on a given orientifold background can be obtained by processes of brane recombinations. This conclusion seems to hold even in the case of non-supersymmetric BSB models. The process of brane recombination is T-dual to the transmutation of D5 branes into magnetic flux. Considering the effective field theory equations of motion for an Abelian gauge theory on a D9 brane in the presence of a FI term mimicking the {\it vevs} of the D9-D5 states, we have shown that a magnetic field arises as a consequence of the compactness of the internal space. The corresponding solution which describes the D9/D5 to magnetized D9 transition was singular in the origin yielding an infinite instanton number. It is reasonable to expect the singularity to be be removed when higher derivative corrections are taken into account.\\

\end{document}